\documentstyle[12pt]{article}

\def\refs{\leftskip=.3truein\parindent=-.3truein}
\def\unrefs{\leftskip=0.0truein\parindent=20pt}

\title {The 9 Aurigae System\thanks{To be published in:
  {\em Monthly Notices of the Royal Astronomical Society},  1993}}

\author {K. Krisciunas\footnotemark[1] \and C. Aspin\footnotemark[1]
\and T. R. Geballe\footnotemark[1] \and H. Akazawa\footnotemark[2]
\and C. F. Claver\footnotemark[3] \and E. F. Guinan\footnotemark[4]
\and H. J. Landis\footnotemark[5] \and K. D. Luedeke\footnotemark[6]
\and N. Ohkura\footnotemark[7] \and O. Ohshima\footnotemark[8]
\and D. R. Skillman\footnotemark[9] }

\date {Accepted 9 February 1993}

\begin{document}
\maketitle

\vspace{5 mm}

\refs

{\em $^{\ast}$ Joint Astronomy Centre, 660 N. A'ohoku Place, University
   Park, Hilo, Hawaii 96720 USA}

{\em $^{\dag}$ Funaho 102, Asakochi, Okayama, 710-02 Japan}

{\em $^{\ddag}$ University of Texas, Department of Astronomy, Austin,
   TX 78712-1083 USA}

{\em $^{\S}$ Villanova University, Department of Astronomy, Villanova,
   PA 19085 USA}

{\em $^{\P}$ 303 Saddle Creek Drive, McDonough, GA 30253-6585 USA}

{\em $^{\parallel}$ 9624 Giddings Ave. NE, Albuquerque, NM 97109 USA}

{\em $^{\ast\ast}$ Senoo 25, Okayama, 702 Japan}

{\em $^{\dag\dag}$ 3-10-15 Tamashima, Kurashiki, Okayama, 713 Japan}

{\em $^{\ddag\ddag}$ 9517 Washington Avenue, Laurel, MD 20723 USA}

\unrefs

\begin{abstract}
The F0 V star 9 Aur A exhibits an irregular variability
of amplitude $\approx $0.1 magnitude at optical wavelengths.  The
variations are too slow for it to be a $\delta$ Scuti-type star.
There is no evidence for a close, interacting companion or
ring of dust, either from infrared, ultraviolet, or speckle
data.  The photometric variability of 9 Aur A is similar to
two other early F dwarf stars: $\gamma$ Doradus and HD 96008.  9 Aur
B appears to be an M dwarf, 9 Aur C is an early- to mid-K
dwarf star, and 9 Aur E, if it is a member of the system,
probably is a normal white dwarf. 9 Aur D is most likely an
unrelated and distant K giant.

\end{abstract}

Key words: Stars, double and multiple - Photometry, 9 Aur -
Stars, variable

{\bf 1. Introduction }

9 Aurigae (= BS 1637 = HD 32537 = $\beta$ 1046 = SAO 25019 =
Gliese 187.2 = ADS 3675 = IRAS 0527+5131) is a multiple star
system whose V $\approx$ 5.0 magnitude primary is of spectral type F0
V.  The distance modulus of the primary is m - M = 1.55 $\pm $ 0.6
(probable error), adopting Gliese's (1969) parallax of 0.049
$\pm$ 0.013 (p.e) arcsec. Wagman (1967) found a parallax of 0.057
$\pm$ 0.005 (p.e.) arcsec.

Krisciunas \& Guinan (1990) found 9 Aur A to be variable
at optical wavelengths by about 0.1 magnitude. Guinan's data,
from 52 nights of observations at Mt. Hopkins using an
automatic photoelectric telescope (APT) from 1989 November to
1990 March, indicated a range of V = 4.96 to 5.07.
Photometry from 1987 January to 1990 April indicated a period
of about 36-39 days (Krisciunas \& Guinan 1990).  However,
subsequent photometry (Krisciunas et al. 1991), while
exhibiting a similar range of brightness, did not bear out
the previously observed periodicity (see Fig. 1).

Given its position in the Hertzsprung-Russell Diagram, 9
Aur should not be variable.  It lies outside the cool edge of
the instability strip (Breger 1979), so it should not
pulsate.  Its light curve from 1987 to 1990 was not
indicative of an eclipsing binary.

This paper investigates possible causes of the
variability in 9 Aur A and proposes rough classifications for
the other fainter components of the 9 Aur system.

{\bf 2. Variability of the Primary: Possible Causes and Recent
Observations}

{\em 2.1  A close companion ?}

Abt (1965) initially reported that 9 Aur A is a
spectroscopic binary with an orbital period of 391.7 days.
However, subsequently Abt \& Levy (1974) ``could not confirm
the previous orbit" and concluded that its radial velocity
was constant, i.e. that the primary had no close companion.
Hartkopf \& McAlister (1984) have made speckle observations of
9 Aur A in order to search for a close stellar companion.
They found nothing to a limit in angular separation of »0.03
arcsec, but would have been unable to detect a close
companion if it were more than 2.5 magnitudes fainter than
the primary (Worley 1992, private communication).

A companion which is considerably hotter or cooler than
the primary could be detected by infrared or ultraviolet
observations.  Near infrared photometry obtained by Geballe
and Aspin, using the United Kingdom Infrared Telescope, is
combined with optical photometry (Hoffleit \& Jaschek 1982) in
Fig. 2. The data are well fitted by a single Planck function
with T = 7179 K.  Simon (1992, private communication) has
obtained an IUE (1100-2200 \AA \ spectrum of 9 Aur, finding no
evidence for a close, hot companion.  Thus, on the basis of
infrared, ultraviolet, and speckle data, there is no evidence
for a close companion of 9 Aur A.

{\em 2.2 Circumstellar material ?}

Could eclipses of 9 Aur A caused by a lumpy ring of dust
around the star be responsible for its variability?  To do so
and be consistent with the observed variability, one might
expect the ring to have a rotational period of $\approx$ 40 days.
Given Kepler's Third Law and the temperature and mass of the
F0 V primary (1.38 M$_{\odot}$
(Takeda 1984) to 1.7 M$_{\odot}$ (Allen 1973, p.
209)), such a ring of dust would be heated to $\approx$ 800 K and
would produce an infrared excess at 3-4 microns.  This is
ruled out by the new infrared photometric data.

9 Aur was detected by IRAS at 12 microns, but not at 25,
60, or 100 microns. The 12 micron flux given in the IRAS
catalogue (1988) is 0.95 Jy, with an uncertainty of $\approx$ 0.08 Jy.
The position angle of the 0.76 by 4.6 arcmin beam (Fullmer \&
Lonsdale 1989) was 81 deg; thus the B component (r $\approx$ 5 arcsec,
position angle $\approx$ 82 deg), the C component (r = 90 arcsec, PA
= 61 deg), and possibly the D component would have been
included in the IRAS beam.  Assuming that the B component is
an M2 dwarf and that the C component is a K5 dwarf (see
below), the 12 micron fluxes of the A, B, and C components
are $\approx$ 0.67, 0.02, and 0.10 Jy, respectively.  The
contribution from the D component is expected to be
negligible (see below). The sum of these fluxes, 0.79 Jy, is
marginally significantly less than the measured IRAS flux.
Thus, there is little evidence of a significant infrared
excess in 9 Aur system and it seems unlikely that variable
obscuration could be the cause of the variability of the
primary.

{\em 2.3 Rotational modulation of surface features ? }

Recently, Abt obtained a reliable value of v sin {\em i} = 20
km/sec for 9 Aur (see Krisciunas et al. 1991).   (The value
of 14 km/sec in {\em The Bright Star Catalogue} (Hoffleit \& Jaschek
1982) is probably based on Huang (1953) and is not
definitive.)  Takeda (1984) gives a radius of 1.95 $\pm$ 0.50 R$_{\odot}$
for the star, based on its position in the HR Diagram.  This
would indicate a rotational period of 4.93 $\pm$ 1.26 days times
sin {\em i}.  Thus the optical variability observed from 1987 to
1990 cannot be attributed to a rotational effect.  One
possibility is that we are observing 9 Aur A roughly pole-on,
and rotational modulation of bright or dark spots would be
seen only if such surface features were near the star's
equator.  Otherwise, the observed variability would be on
time scales corresponding to the growth and decay of the
surface features.

{\em 2.4 Pulsations ?}

Taking the reported radial velocities at face value (Abt
\& Levy 1974; Takeda 1984; Duquennoy et al. 1991), it is
conceivable that the radial velocity of 9 Aur varies from
-2.1 to +3.5 km/sec. If 9 Aur were a $\delta$ Scuti star, its
principal period of pulsation would be some tens of minutes.
Krisciunas et al. (1991) sought evidence for short term
pulsations of 9 Aur, based on photometry lasting up to 5.6
hours per night and radial velocity measurements (on one
occasion 28 radial velocities were obtained over a span of 2
hours).  No evidence of pulsations was found from the
photometry or radial velocities.  From December 1990 to March
1991 the observed range of brightness was V = 4.93 to 5.00
from the nightly means, but real, shorter term variations
were observed on several individual nights (see Fig. 3),
which were linear changes of brightness of about 0.01 mag/hr.
Skillman's data of 1991 February 1-17 UT, which are contained
in IAU File 238 of Unpublished Observations of Variable Stars
(see Breger et al. 1990), indicate possible periods of 0.54
and 1.37 days, but these may just be artifacts of aliasing.
If such periods exist, they could only be substantiated by
data taken over several days by telescopes distributed around
the Earth in longitude.

{\em 2.5 Recent photometry}

A recent optical campaign by observers situated from the
eastern United States to Japan was carried out in 1992 from
January 31 to February 10 UT. Altogether 237 useful
differential V-band observations were obtained of 9 Aur vs.
BS 1561. Table I lists the parameters of the observations
including the number of differential measures of 9 Aur and
the typical internal error of individual differential
measures (primarily from 120 observations of BS 1568 vs. BS
1561). Figure 4 displays the nightly means during the
observing period.  The complete data set can be found in IAU
File 244 (see Breger et al. 1990).  From the nightly means
the observations show variations of up to 0.057 mag.  There
is no evidence in these data for the previously reported 36-
39 day variations or for linear variations of 0.01 mag/hour.

To search for periodicities the above data were analyzed
by means of the Lomb-Scargle algorithm (see Press \& Teukolsky
1988).  We used an oversampling factor of 4.  Figure 5 shows
the V-band power spectrum up to frequencies of half the
Nyquist frequency.  No significant peaks are to be found from
0.5 to 2.0 times the Nyquist frequency.  The highest peak of
the resultant power spectrum corresponds to a period of 2.725
days, with a false alarm probability of $10^{-20}$.  In Fig. 6 we
show the data folded by a period corresponding to the most
significant peak in the power spectrum.  Other peaks show up
in the power spectrum, such as a 1.277 day peak with a false
alarm probability of $10^{-18}$.  Peaks near 1, 2, and 3
cycles/day may be artifacts of aliasing, since we did not
have round-the-clock coverage.

Frequency analysis of the check star (BS 1568) vs. BS
1561 data gave no significant peaks in the power spectrum
corresponding to the just-mentioned periods, and no peaks
with false alarm probabilities less than about 0.03. This is
further evidence that both the comparison star and check star
are constant, and that 9 Aur is a bona fide variable.

Because of the large scatter of data on a given night
for a given observer compared to the variations of the
nightly means, we do not place much trust in the permanence
of the peaks in the power spectrum.  Still, the data provide
evidence that variations are taking place on short time
scales, suggesting rotational modulation of surface spots as
a possible cause.  If 2.725 days were the true rotational
period of 9 Aur A, equating it to the period derived from the
rotational velocity (see \S 2.3) implies an inclination angle
of $\approx$ 34 deg.

{\em 2.6 Colors, Abundances}

Photometry given by Eggen (1963), Hoffleit \& Jaschek
(1982), and obtained by us indicate that 9 Aur A has a B-V
color in the range 0.30-0.35. Its measured U-B color ranges
between +0.017 and -0.04, and perhaps as low as -0.13.  (We
do not imply that the colors are variable - just that
different observers obtained data that are systematically
different.)  According to the data in Hoffleit \& Jaschek, F0
V stars have $<$B-V$>$ = 0.28 and $<$U-B$>$ = +0.07. Thus, 9 Aur A
has a B-V color several hundredths of a magnitude redder than
the ``average" F0 dwarf star, and an apparent UV excess of
0.05-0.11 (and perhaps 0.20) mag.  If this $\approx$ 0.1 mag excess
were attributed an underabundance of metals, the implied
[Fe/H] would be approximately -0.4 (Carney 1979).  However, 9
Aur A has essentially solar abundances.  Provost \& van't
Veer-Menneret (1969), Bell (1971), and Cayrel de Strobel et
al. (1980) give [Fe/H] = +0.01, +0.03, and +0.10,
respectively.

{\bf 3. The Fainter Components of the 9 Aur System}

Table II summarizes the observed properties of the other
stars in the 9 Aur system. Additional measurements and
comments on these stars are given below.  Proper motions and
parallaxes have not been reported for the fainter components.
We assume the same distance modulus for them as for 9 Aur A.

{\em 3.1 The B Component }

A low resolution K-band spectrum of this star was
obtained on 1992 January 7 UT at UKIRT, using the facility
spectrometer, CGS4 (Mountain et al. 1990). No spectral
features were clearly seen, but a weak CO band was marginally
detected.  A similar spectrum of the primary revealed only a
Br g absorption line, as expected for an F star.

Koornneef (1983) gives V-K = 3.75, J-K = 0.89, and H-K =
0.21 for an M2 dwarf star.  Both the colors of the B
component and its K-band spectrum are consistent with this
classification. The optical photometry of the primary
included the light of the B component. As the magnitude of
the variations in the photometry corresponds to variations of
Å 4 magnitudes in the B component, it is unlikely that the B
component is responsible for them.

{\em 3.2 The C Component }

  Griffin (1992, private communication) has measured the
radial velocities of the C star and the primary.  From the
preliminary data the two radial velocities differed by only
3.8 km/sec, suggesting that the C component is in fact a
companion.  The visual magnitude and colors (Table II) imply
that the C component is an early-to-middle K dwarf (see Allen
1973, p. 106)

{\em 3.3 The D Component }

Spectroscopy of 9 Aur D by P. Hendry and J. Thomson on
1992 May 5, covering the interval 6330-6785 \AA, shows numerous
metallic lines in addition to H a absorption (Bolton 1992,
private communication). A radial velocity of +13.1 $\pm$  1.0
km/sec was derived by comparison to HD 107328, whose radial
velocity was assumed to be +35.7 km/sec.  The radial velocity
suggests that the D component is not a member of the 9 Aur
system.  Other indicators of this are its visual magnitude
and B-V color, which place it below the main sequence yet
above the white dwarf cooling track, and are not consistent
with any known type of star if it is at the distance of 9 Aur
A.  Therefore, it appears most likely that 9 Aur D is a
distant K giant.

{\em 3.4 The E Component}

The implied absolute magnitude and color of this object
are similar to those of some white dwarfs (McCook \& Sion
1987).  The E star may be a white dwarf companion of 9 Aur,
but it could also be a more distant main sequence star.
Spectroscopic observations are needed to test this.

{\bf 4. Discussion}

The motivation of this work was to find an explanation
for the unexpected and unexplained photometric variability of
9 Aur A, which, apart from its variability, appears to be a
normal F0 V star.  This variability could be interpreted as a
rotational modulation of activity in the atmosphere of the
star.  If the star is viewed close to pole-on, as is
suggested in \S 2.5, the time scale for variability might be
the time scale for the growth and decay of disturbances,
rather than the rotational period.  It is also possible that
(non-radial) pulsations are the cause of the variation;
further radial velocity data are required to test this.

9 Aur A has few spectroscopic anomalies.  However,
Coupry \& Burkhart (1992) found that the Fe I 6678 \AA \ and Ca I
6717 \AA \ absorption lines were asymmetric, with blue wings.
They and Provost \& van't Veer-Menneret (1969) note the star's
high microturbulence velocity.  It may be relevant to
determine if these anomalies are variable.

As 9 Aur A appears to be intrinsically variable, it is
of interest to know if there are other similar stars.  Two
early F stars with normal UBV colors and unexplained
variability similar to 9 Aur are $\gamma$ Dor (Cousins 1992) and HD
96008 (Lampens 1987; Matthews 1990).  $\gamma$ Dor has two periods,
each near 3/4 day, which suggests non-radial stellar
pulsations.  HD 96008 has a period of 0.31 days.  Matthews
notes that the variability of HD 96008 is not satisfactorily
explained by rotation, given the implied rotational speed of
Å330 km/sec, but that $\delta$ Scuti-type pulsation is not
satisfactory either due to the combination of period and
luminosity (i.e. the observed period is too long for a normal
$\delta$ Scuti star).  In short, 9 Aur, $\gamma$ Dor, and HD 96008 are three
similar stars whose variability still remains unexplained.

Other F stars with unusual colors similar to those of 9
Aur A are BS 4825/6 and BS 4914.  One of the former pair is
variable with a range of 0.03 mag (Hoffleit \& Jaschek 1982).
BS 4914 is the companion of $\alpha^{2}$ CVn, a prototypical magnetic
rotator. Photometry of other early F dwarf stars with B-V
redder than ``average", U-B bluer than ``average", and moderate
rotational velocities may reveal other small amplitude
variables.  Examples are BS 5074, 5075, and 6600.  Two stars
of similar color with high rotational velocities, which
should be investigated, are BS 463 and BS 7887.

Because the ``true" period(s) of variation for 9 Aur have
not been determined, further coordinated photometry by
observers situated around the globe would be useful.
Unfortunately, 9 Aur is too bright to be a target of the
Whole Earth Telescope (Nather \& Winget 1992). More extensive
and highly accurate radial velocity measurements also are
needed.  In addition to further observations, we suggest that
greater scrutiny of stellar atmosphere models of early F
dwarf stars is warranted.

\begin {center}
{\bf Summary}
\end {center}

The cause(s) of the photometric variations of 9 Aur A
remain uncertain, although a close companion, variable
obscuration, and contamination of the photometry from other
members of the 9 Aur system are ruled out.  Investigation of
the other components of 9 Aur have led to the following
conclusions: The B component appears to be an early-M dwarf
and the C component is an early- to mid-K dwarf, both within
the system; the E component may be a white dwarf within the
system; the D component appears to be a distant K giant.

\begin {center}
  {\bf Acknowledgments}
\end {center}

The UKIRT data discussed here were obtained as part of
the UKIRT Service Observing and Astronomer in Charge
Discretionary Time programs.  The opportunity to obtain such
data is greatly appreciated.  This paper is based in part on
information from the SIMBAD data retrieval system, data base
of the Strasbourg, France, Astronomical Data Center.  KK
thanks John Gathright and the University of Hawaii, Hilo, for
the loan of an Optec SSP-3 photometer; he also thanks the
University of Hawaii, Manoa, for telescope time on their No.
1 24-inch telescope at Mauna Kea.  We thank Roger Griffin,
Tom Bolton, Paul Hendry, Jim Thomson, and Ted Simon for their
efforts in securing preliminary data discussed here.  We
thank Tim Hawarden, Alan Batten, Wulff Heintz, Charles
Worley, Ed Nather, and Chris Sneden for useful discussions.
Finally, we thank the referee, Luis Balona, for bringing to
our attention the papers on $\gamma$ Doradus and HD 96008.

\begin {center}
 {\bf References}
\end {center}

\refs

Abt, H. A., 1965, ApJS, 11, 429

Abt, H. A., Levy, S. G., 1974, ApJ, 188, 291

Allen, C. W., 1973, Astrophysical Quantities, 3rd ed.
   Athlone, London

Bell, R. A, 1971, MNRAS, 155, 65

Breger, M., 1979, PASP, 91, 5

Breger, M., Jaschek, C., Dubois, P., 1990, IBVS, No. 3422

Carney, B. W., 1979, ApJ, 233, 211

Cayrel de Strobel, G., Bentoila, C., Hauck, B., Curchod,
   A., 1980, A\&AS, 41, 405

Coupry, M. F., Burkhart, C., 1992, A\&AS, 95, 41

Cousins, A. W. J., 1992, Observatory, 112, 53

Duquennoy, A., Mayor, M., Halbwachs, J.-L., 1991, A\&AS, 88,
   281

Eggen, O. J., 1963, AJ, 68, 483

Fullmer, L., Lonsdale, C., 1989, Catalogues Galaxies and
   Quasars Observed in the IRAS Survey, version 2.
   IPAC and Caltech, Pasadena, CA, p. II-1

Gliese, W., 1969, Ver\"{o}ffentlichungen des Astron. Rechen-
   Inst., No. 22

Hartkopf, W. I., McAlister, H. A., 1984, PASP, 96, 105

Huang, S.-S., 1953, ApJ, 118, 285

Hoffleit, D., Jaschek, C., 1982,  The Bright Star
   Catalogue, 4th ed. Yale Univ. Observatory, New Haven, CT

Jeffers, H. M., van den Bos, W. H., Greeby, F. M., 1963,
   Index Catalogue of Visual Double Stars, 1961.0.
   Lick Observatory, Mt. Hamilton, CA

Joint IRAS Science Working Group, 1988, Infrared Astronomical
   Satellite (IRAS). Catalogs and Atlases, vol. 2.
   NASA, Washington, D. C., p. 190

Koornneef, J., 1983, A\&A, 128, 84

Krisciunas, K., Guinan, E., 1990, IBVS, No. 3511

Krisciunas, K., Guinan, E. F., Skillman, D. R., Abt, H. A.,
  1991, IBVS, No. 3672

Lampens, P., 1987, A\&A, 172, 173

Matthews, J. M., 1990, A\&A, 229, 452

McCook, G. P., Sion, E. M., 1987, ApJS, 65, 603

Mountain, C. M., Robertson, D., Lee, T. J., Wade, R. 1990.
   SPIE 1235. Instrumentation in Astronomy VII, p. 25.

Nather, R. E., Winget, D. E., 1992,  Sky \& Tel., 83, 374

Press. W. H., Teukolsky, S. A., 1988,  Computers in
   Physics, 2, No. 6 (Nov/Dec), 77

Provost, J., van't Veer-Menneret, C., 1969, A\&A, 2, 218

Takeda, T., 1984, PASJ, 36, 149

Tokunaga, A., 1986, The NASA Infrared Telescope Facility
   Photometry Manual. University of Hawaii, Institute for
   Astronomy, Honolulu

Wagman, N. E., 1967, AJ, 72, 957

\unrefs

\begin {center}
  {\bf Note Added in Press}
\end {center}

A fourth member of this class of early F-type ``variables without
a cause" is HD 164615 (Abt et al. 1983, {\em MNRAS} {\bf 272}, 196;
Rucinski 1985, {\em PASP} {\bf 97}, 657), which is classed as
F2 IV-V.  Its photometric period was found to be $\approx$ 0.815 days.
Abt et al. obtained 13 radial velocities which ranged 7.4 km/sec, but
on the basis of the relatively large internal errors of the individual
measurements concluded that the radial velocity was constant.

Griffin (1993, private communication) obtained 20 radial velocities of
9 Aur in February and March of 1993.  Further analysis must be
carried out on these data, but from the preliminary data a range of
$\approx$ 6 km/sec is indicated, with a period of less than 3 days.
Thus it seems that 9 Aur exhibits
radial velocity variations on a time scale comparable to that of the
photometric variations.

\vspace {10 mm}
\begin {center}
{\bf Figure Captions}
\end {center}

Fig. 1 - Nightly means of differential photometry of 9
Aur vs. BS 1561.  Circles: data by Guinan with a 25-cm
automatic photoelectric telescope (APT) at Mt. Hopkins.
Squares: data by Skillman, using a 32-cm APT.  Guinan's means
are the average of 18 to 54 points per night.  Skillman's
means are the average of 5 to 35 points per night.  On 5 of
these 14 nights 9 Aur showed evidence of linear variations on
the order of 0.01 mag/hr.

Fig. 2 - The multi-wavelength flux of 9 Aur can be
fitted by a Planck curve with T = 7179 K.  From left to
right, the filters are U, B, V, J, H, K, L, L', and narrow-
band M.  Squares: optical data given by Hoffleit \& Jaschek
(1982).  Dots: photometry of Geballe, using UKIRT and the
single channel photometer UKT9.  Triangles: data by Aspin,
using UKIRT and the facility infrared array camera IRCAM.
All the infrared photometry excludes the B component.  Except
for U and B, the conversion of magnitudes to flux in Jy was
done according to the $\alpha$ Lyr flux given by Tokunaga (1986).  U
= 0 and B = 0 were taken to correspond to 2000 and 3700 Jy,
respectively.  The U-band point was excluded from the least-
squares fit.

Fig. 3 - Differential V-band photometry of 9 Aur vs. BS
1561 by Skillman over the course of 5 hr 35 min on 1991 Feb 1
UT. Once the linear trend amounting to 0.01 mag/hr is
removed, there are no significant periodic variations on time
scales of tens of minutes (like a $\delta$ Scuti star).

Fig. 4 - Nightly V-band means of 9 Aur vs. BS 1561
obtained from 1992 Jan 31 to Feb 10 UT.  X's: data by
Krisciunas at 2800-m elevation of Mauna Kea. Dots: data by
Krisciunas at 4200-m Mauna Kea summit, using UH 24-inch
telescope.  Filled triangle: data by Luedeke (average of 27
points).  +: data by Landis (average of 12 points).  Open
circle: data by Ohshima (average of 63 points).  Open
squares: data by Akazawa.  Open triangles: data by Ohkura.

Fig. 5 - Power spectrum of 237 individual V-band
differential magnitudes of 9 Aur vs. BS 1561, using the Lomb-
Scargle algorithm (Press \& Teukolsky 1988).

Fig. 6 - Differential photometry of 9 Aur vs. BS 1561.
The symbols are the same as in Fig. 4.  The data are folded
by a period corresponding to the most significant peak in the
power spectrum.  Systematic errors between different
observers would amount to 0.01 mag or less.

\end{document}